\documentclass[10pt, a4paper]{article}

\usepackage[utf8]{inputenc}
\usepackage{amsmath}
\usepackage[space]{cite}
\usepackage[T1]{fontenc}
\usepackage{babel}
\usepackage{graphicx}
\usepackage[font=small,labelfont=bf]{caption}

\title{Realistic gossip in Trust Game on networks: the GODS model}

\author{
  Jan Majewski\textsuperscript{1*}
  \\
  Francesca Giardini\textsuperscript{2}\\
}

\date{\footnotesize\textsuperscript{\textbf{1}}University of Warsaw, Faculty of Sociology, Department of Digital Sociology, ul. Karowa 18, Warsaw, 03-247, Poland\\ \textsuperscript{\textbf{2}}University of Groningen, Department of Sociology and ICS, Grote Rozenstraat, Groningen, 31 - 9712 TG, Netherlands\\
\texttt{*jj.majewski2@uw.edu.pl}\\
\normalsize First version: Septermber 4th 2024\\
This version: October 1st 2025}

\begin{document}
\maketitle 

\begin{abstract} Gossip has been shown to be a relatively efficient solution to problems of cooperation in reputation-based systems of exchange, but many studies don’t conceptualize gossiping in a realistic way, often assuming near-perfect information or broadcast-like dynamics of its spread. To solve this problem, we developed an agent-based model that pairs realistic gossip processes with different variants of Trust Game. The results show that cooperators suffer when local interactions govern spread of gossip, because they cannot discriminate against defectors. Realistic gossiping increases the overall amount of resources, but is more likely to promote defection. Moreover, even partner selection through dynamic networks can lead to high payoff inequalities among agent types. Cooperators face a choice between outcompeting defectors and overall growth. By blending direct and indirect reciprocity with reputations we show that gossiping increases the efficiency of cooperation by an order of magnitude. \end{abstract}
\begin{flushleft}
\end{flushleft} 

\sloppy

\section{Introduction}

The problem of cooperation\cite{axelrod1981evolution} is a core scientific problem that crosses domains as diverse as the social and the natural sciences. Among other solutions, gossip-based reputation systems \cite{giardini_gossip_2019, milinski2019gossip} that include both direct and indirect reciprocity\cite{schmid_unified_2021}, have been advocated as one of the most frequent in human societies, and one of the most successful across different domains \cite{szamado2021language}. For reputations to support cooperation, they must be reliable, i.e., accurate in portraying the main features of someone's interaction partner, and available to many other actors. Although the experimental and simulation literature on the interaction between gossip and cooperation is extensive, many authors lament a lack of realism in the implementation of gossip in the lab and in agent-based simulations \cite{fonseca_is_2021, righi_gossip_2022, giardini_four_2022, correia2025evolution}. Due to the limitations imposed by the laboratory space, where intimacy and confidentiality cannot be guaranteed, most papers model gossip as a centrally organized exchange of messages among players\cite{milinski_reputation_2016}. Others allow participants to exchange messages with selected partners \cite{giardini_gossip_2021}, but the question of the realism of these studies, i.e., the extent to which they can faithfully reproduce the phenomenon that they want to study, remains open.

Human groups are characterized by high levels of connectedness. This means that people interact along specific, but flexible social structures, with many possible network configurations. The relevance and impact of social relationships on individual behaviors is widely accepted, but it is seldom considered when studying gossip \cite{takacs_networks_2021, giardini_four_2022} and reputation \cite{wittek2023evolution} in a non-stylized way\cite{okada_cooperation_2021}.

Modeling a realistic process of gossiping on different networks can provide useful insights into the workings of reputation-based cooperation applied to trust games. Trust is essentially an expectation of a person A about a person B (A judges the trustworthiness of B) with regard to behavior X in context Y at time T \cite{uslaner_measuring_2017}. Trus­t between partners (as opposed to generalized trust) is built as a consequence of repeated exchanges \cite{kollock_emergence_1994} and allows for creation of alliances capable of solving problems that are beyond the power of individuals\cite{coleman_foundations_2000}. Trust Game directly measures this relationship, as each give-and-return interaction reveals what kind of partner one is dealing with (as opposed to Public Goods Game where free-riders can be detected on a group level) in an orderly way (as opposed to Prisoner’s Dilemma where each player acts on the basis of double contingency).

We choose a Trust Game because, among social dilemma games used to simplify different types of human interactions\cite{van_dijk_experimental_2021}, the Trust Game (TG) seems to be an especially fair portrayal of many real-life situations in which there is asymmetry between the players. The Trust Game (TG) is a two-person game in which players (a trustor who moves first and a trustee moving second) have an opportunity to exchange resources and increase their total amount, but are tempted by egoistic incentives to keep whatever they get to themselves (original formulation as investment game\cite{berg1995trust}; computational extensions\cite{diekmann2005evolution, kumar_evolution_2020}. Rational choice theory predicts that, in such a scenario, a reasonable actor doesn’t cooperate, but experimental evidence suggests that the opposite is true\cite{johnson_trust_2011}: people have a general tendency to share with others. At the same time, this initial prosociality can plummet as subsequent rounds are played\cite{milinski_reputation_2002}, especially if each round is played with another partner, eliminating direct reciprocity. The TG has been extensively used in laboratory experiments\cite{johnson_trust_2011}, and in simulation models\cite{chen_network-based_2015} to investigate prosocial behavior in dyadic interactions. This experimental paradigm has been used also in applied contexts, such as interfirm relations \cite{coleman_foundations_2000, greif_contract_1993}, online markets shopping\cite{diekmann_trust_2019, jiao_reputation_2021} and intra-organizational dynamics\cite{burt_brokerage_2005}.

The possibility of being talked about, and especially the talking itself, introduce both internal and external pressures pushing individuals towards cooperation in TG\cite{boero_reputational_2009, feinberg_virtues_2012, fehr2019gossip, fonseca_will_2018, peters_truth_2020}. Gaining good reputations becomes a concern for its own sake when players regularly exchange gossip\cite{feinberg_gossip_2014}, and it even leads to subsequent cooperation among defectors, since it is the only way to get back in the game, especially when ostracism or partner selection\cite{giardini_gossip_2021} are present. This last aspect, often introduced in experiments, simulations and empirical studies by the means of dynamic networks\cite{gallo_effects_2015, takacs_networks_2021, nee2023theory}, draws our attention to adaptive pressures of different situations on both cooperators and defectors – even prosocial individuals can become defectors (or very cautious cooperators at best) if they are exploited too much, while defectors (if motivated enough) have potential to become good partners.

However, both the experimental and the modeling literature about gossiping in TG have limitations in terms of realism. One of the largest problems with literature on gossip-induced cooperation has been lack of realistic interactions driving gossip spread. The assumption that gossip is omnipresent and available to everyone has been taken for granted in many studies\cite{emler1994gossip, dunbar_gossip_2004, ohtsuki_reputation_2015, yucel2021being}. In real life gossip is present in many different social milieus often, but not all the time\cite{dores2021gossip}. In everyday conversations positive and neutral gossip may be more popular than negative\cite{robbins_who_2020, dores2021gossip}, but negative gossip is most often recalled in vignette studies\cite{martinescu_what_2022} and seems to make the most impact. It seems reasonable to say that it is performed by everyone, but probably not all the time and definitely not with everyone. For instance, Estévez and Takács\cite{estevez_brokering_2022} discovered that only 1\% of all possible sender-receiver-target triads have been found in real gossip networks within organizations. These findings suggest high selectivity of gossip partners, localized transfer of information and its ephemeral character.

Another feature of gossip models that is far from reality is the fact that gossip is freely available and abundant. The quantity of gossip has been demonstrated to be a factor determining the emergence of cooperation\cite{giardini_evolution_2016}, but how much gossip people actually exchange in reality is a still unclear\cite{dores2021gossip}. In reality, it is unusual to receive so much information and to act on gossip we just heard; individuals rather pick up a few pieces of gossip about numerous people from very few and selected sources and over longer periods. On the other hand, in experiments with voluntary communication we observe a rather limited amount of gossip interactions: one in four\cite{fehr2019gossip} and half\cite{samu_scarce_2020} of gossip statements are actually emitted. When gossip is voluntary, it usually occurs early on in the game and supports cooperation for some time (contributions change according to gossip’s valence). There is a limit to this effect and introducing more gossiping within one round (e.g. for purposes of cross-checking or social bonding\cite{samu_evaluating_2021}) can lead to an overall decrease in cooperation levels, so it may not necessarily be the case that huge quantities of reputational information are needed for human societies to operate, but that is a topic for another research.

The aim of this paper is to model gossip in a TG in such a way to address all these limitations and to verify if it still supports cooperation. Our first contribution is modeling more realistic gossip interactions. We propose a localized vision of gossiping performed on empirical, signed social networks of affect (friendship-antipathy) gathered for the purposes of studying gossip in organizations\cite{estevez_brokering_2022, estevez_more_2022}. Identity of gossip partners (sender-receiver-target triad), content of gossip (type of information) and its valence (positive vs. negative) have all important consequences for the process of gossip spread. For instance, empirical research indicates that positive gossip is invited by positive relationships within the gossip triad\cite{grosser2010social}, while having a common enemy induces negative talk\cite{ellwardt_co-evolution_2012}. On the other hand, there are certain interpersonal configurations that may completely inhibit gossiping\cite{giardini_silence_2019}. All these complex interrelations create a set of distinguishable situations which can be conceived of as social mechanisms governing transmission of gossip.

The second contribution of this paper is in the modeling of many types of TG-gossip interactions. We consider game, reputational and network scope of this dilemma. Social simulations should always explore plausible model designs not only for strictly parametric elements (like in OFAT sensitivity analysis), but also for different procedures\cite{borgonovo_sensitivity_2022} to foster more understanding. This has been implemented here as three distinct sets of behavioral rules for making decisions in the Trust Game on the basis of reputations. In addition, we tested for different combinations of direct and indirect reciprocity\cite{schmid_unified_2021} by manipulating the amount of TG interactions per time interval (see SI). This allows us to see whether gossiping has a potential for supporting cooperation not only when talking is scarce, but also when direct experiences are limited.

\section{Methods}

This work rests on the agent-based model called GODS: Gossip-Oriented Dilemma Simulator (see SI and ODD for further details). The purpose of the model is to bring agent behavior closer to real-life interactions (see KISS-EROS discussion\cite{jager2017enhancing}) without losing generalizability of conclusions.

We set a population of N agents (from 16 to 36) connected by an empirical social network (see ODD for details on datasets and topologies). The agents alternate playing a round of TG and gossiping on the social network. Depending on who the agent is and who it perceives to play against, it decides to cooperate or defect. The direct experiences of agents are stored in matrix $I$, which serves as an information base for gossiping (matrix $R$). Agents are initialized with 20 units of resources, a strategy (a C or D type) and access to reputation knowledge – formalized as a square N-by-N image matrix $I$ which stores past behavior of agents in TG, while the square N-by-N reputation matrix $R$ stores gossip information received from third parties.

\subsubsection{Trust Game}
Trust Game is a sequential, two-player game in which the Trustor is endowed with an amount of resources and decides how much of it to transfer to the Trustee. The chosen amount is then multiplied by 3 and transferred to the Trustee, who chooses if they want to return some of the received amount. There are four possible outcomes of TG interaction (see Fig. 2 in ODD). During every TG round each agent is selected to play once as Trustor and once as Trustee. Cooperation is costly in terms of resources, but is rewarded by a good reputation, while defection pays in resources, but leaves a bad reputation. TG outcomes shape reputations directly (image bound by -1 and 1). This is implemented as an image matrix $I$ recording direct experiences of each trustor and trustee. We assume that both Cs and Ds act reasonably in assigning reputations (see mathematical treatment of TG\cite{masuda_coevolution_2012}). Empirical evidence suggests that people show a tendency towards inheritance of reputation scores\cite{samu_scarce_2020}, which means that reputations are influenced by both immediate action and an earlier reputation score. For that reason, we propose a simplifying assumption that reputation-formation follows well-known dynamics of assimilative influence\cite{flache_models_2017}, but here it is even simpler, as each new value is just a modification of an old one – good actions make it better and vice versa. That is implemented by entry $I_{ij}$ (agent i’s perception of agent j in matrix $I$) at time t which is a sum of their standing at time t-1 and current action:

\begin{equation}
  I_{ij_{\ t+1}} =
    \begin{cases}
      I_{ij_{\ t}} + 0.1  & \text{if $j$ cooperates with $i$ at time $t$}\\
      I_{ij_{\ t}} - 0.1  & \text{if $j$ defects against $i$ at time $t$}\\
      I_{ij_{\ t}} & \text{otherwise}
    \end{cases} 
\end{equation}

Agents' decisions in the TG are based on a partner's reputation and on one of the action rules. Players generally follow their type, but Cs defect against players with bad reputation (partner’s reputation is below cooperation threshold). Across variants, players of type D free-ride on others (with exceptions, as in Fig. 3 BIII in ODD). Overall, the baseline scenario does not allow for any adaptations of strategies, as agents do not analyze interactional history of their partners (social norms) and use only image and reputation to make their decisions.

\subsubsection{Agents' behavioral repertoires in the game}
There are three versions of TG action rules (Fig. 3 in ODD). In the first one, cooperators (C) condition their decision on a partner’s reputation (cooperate with partner perceived by them as other Cs and defect against agents perceived as Ds), while defectors (D) always cheat (ALLD). The second strategy supplements the first one with ‘blind reciprocity’. It means that Cs condition their decisions on reputation, but always reciprocate cooperation (regardless of their perception of partner), while Ds continue with ALLD. The third set of action rules adds a leniency period for each agent. That makes Cs' actions conditional on their partners' reputation, but they always reciprocate cooperation and forgive each D for a fixed number of times. Ds have analogous period for cooperation with other agents (regardless of their standing), but once this period runs out, they ALLD. This last feature allows agents to gain good reputations first, and only then proceed with defections.

\subsubsection{TG interaction regime}
We implemented different partner selection mechanisms in TG: well-mixed (random matching of partners), static network (agents play mostly with their neighbors) and dynamic networks (agents choose who they want to play with based on reputations). In all three variants each agent is always chosen for playing as a Trustor exactly once per round and up to N-1 times as Trustee. 

Well-mixed variant makes each agent an equally likely TG interaction partner in role of Trustee, so it is possible to be everyone’s partner or not be chosen as Trustee at all, but after r rounds each agent has the probability not to be chosen at all of $ (1-  \frac{1}{N})^{rN-r}  $, which is small enough to ignore for considering averages from multiple runs. 

TG network is a random network generated by pairing each agent with randomly selected agent d number of times, which creates a network with fixed minimum degree of d and a maximum degree of N-1. Static networks introduce the neighborhood structure (agents connected to Trustor in TG network) and a probability (0.95 in main text) of selecting Trustee from one of those neighbors. Every agent plays once as Trustor per round, so it has a probability of 0.05 of playing with an agent outside the TG network neighborhood (giving each non-neighbor a probability of $0.05\times\frac{1}{N-(\text{degree of Trustor}+1)}$ and giving each neighbor a probability of $0.95\times\frac{1}{\text{degree of Trustor}}$ of playing as a Trustee. In dynamic networks the same mechanism plays out, but is supplemented with possibility of swapping ties in TG network after each round. Every agent uses image matrix (direct experience) to choose the worst partner and drop the tie with them, if they have it. Then this agent can create a tie with an agent they know to be the best partner from matrix $R$ (gossip), if they aren’t yet connected. This process is repeated for all agents after every round is played, so at minimum no ties change (when nobody is connected to the worst partner they had and is connected to the best partner they have heard of) and at maximum 2N ties change (when everyone connects to someone new and drops one of their neighbors without any overlap). To ensure sufficient connectedness of the network (lack of isolates), if an agent has no TG network ties, it links to a random other.

\subsubsection{Time}
Time is discrete and each model step represents one gossip interaction (a dyadic exchange of information). Once per many gossip interactions (frequency set by model parameter) the model executes a round of Trust Game and agents get their resources and form images of their partners. After the single piece of gossip has run its course (frequency set by model parameter), the model sums up a gossip diffusion and starts a new one (see below). Simulation ends after the time has passed and all agents completed their TG rounds.

\subsubsection{Gossip spreading}
We designed and compared two mechanisms for the spreading of gossip: a parallel transmission mechanism of spread, where everyone knows everything about everyone else; and a more localized diffusion process informed by social mechanisms reconstructed from empirical literature.  The former\cite{ohtsuki_reputation_2015} takes all information on past behavior of an agent A and averages it out. Agent A is perceived in the same way by everyone (perfect information) and everyone’s experiences contribute to this perception equally.
Reputations spread by gossip simulated as a series of information diffusions on real social networks. We use publicly available network datasets (see ODD protocol) containing information on emotional charge of ties linking nodes. Agents learn reputations of different gossip targets from each other and use this information to update their perception of them. Formula 2 illustrates the mechanism of updating agents’ reputations after receiving gossip.

\begin{equation}
\resizebox{.9\hsize}{!}{$R_{ij_{\ t+1}} =
    \begin{cases}
       (R_{ij_{\ t}} \times (1 - \omega)) + (I_{sj_{\ t}} \times (\omega)))  & \text{if $i$ receives gossip about $j$ by the time $t$}\\
      R_{ij_{\ t}} & \text{otherwise}
    \end{cases}$}	
\end{equation}

Gossip information is stored in a matrix $R$ independent of image $I$. Incoming information spread by gossip is weighted by $\omega$ factor with the reputation each agent already has about the target. This makes it possible for an agent to have two conflicting sources (for instance, when a direct experience of defection is in contrast with reputation). The resolution of such situations is given by a simple averaging function that sums R and I information weighted by some fraction and its complement to unity accordingly.

Triadic mechanism is implemented as rules governing transmission for the set of all possible situations that can arise within an undirected gossip triad. This type of gossip is spread on empirical, signed social networks of affect (friendship-antipathy). Every node on this network is an agent and every edge (tie) represents a relationship connecting her with another agent. Valence of a tie (sign – positive or negative) determines what type of relationship connects two agents. Whenever an agent receives a piece of gossip, she must decide whether to transmit it to one of her neighbors based on what relationship she (now as a sender) has with her neighbor (receiver) and the target of gossip, as well as the content of gossip (reputation of the target that gives valence of the gossip message – positive/neutral or negative). There are 36 situations (18 gossip triad configurations for two valences of gossip information) each ending with yes or no decision – all choices have been informed by empirical literature (see ODD). Gossip is ephemeral (it has a short lifespan) and people are selective with it\cite{fehr2019gossip}, so we assume that a selection of all reputations (in most scenarios up to 10) can be spread after a single TG round.

\section{Research questions}

Below \ref{tab:1} we present our research questions in the form of specific hypotheses with references to scientific literature and means of verification. The support of cooperation is measured in TG units. The differences in resource distributions among groups allow us to locate not only the dominant variant, but the extent to which one group outcompetes the other (how much richer or poorer are the Cs), as well as the overall inequalities of resource distribution. The latter resonates with larger problem of interplay between inequalities, reputation, cooperation and growth\cite{galor2004physical, melamed_inequality_2022}. Here we focus on specific questions about the influence of availability of information on game behavior.

\begin{table}[ht]
\small{
\centering
\begin{tabular}{p{0.28\textwidth}p{0.22\textwidth}p{0.19\textwidth}p{0.20\textwidth}}
\hline
\textbf{Formulation} & \textbf{Main reference} & \textbf{Key parameters} & \textbf{Indicator}\\
\hline
Hypothesis 1: Parallel transmission of gossip increases cooperators' resources more than triadic transmission of gossip. & Fonseca and Peters 2018 & Gossip mechanism \newline Reputation thresholds \newline Proportion of Ds & The difference of mean resources between Cs and Ds expressed as the number of standard deviations of total resources. \\
\hline
Hypothesis 2: Triadic gossiping will increase resources of population and Cs, compared to parallel transmission. & Tsvetkova 2021 & Gossip mechanism \newline TG interaction \newline Action rules & Mean, standard deviation and absolute difference of resources at the end of the simulation. \newline C-win dummy variable. \\
\hline
Hypothesis 3: Dynamic networks have a positive effect on the performance of cooperators. & Takács et al. 2021 & Well-mixed TG \newline Static networks \newline Dynamic networks & The difference of mean resources between Cs and Ds expressed as the number of standard deviations of total resources. \\
\hline
\end{tabular}
}
\caption{Overview of hypotheses and ways of addressing them.}
\label{tab:1}
\end{table}

The first hypothesis focuses on the problem of information availability. Building on empirical literature of reputation-cooperation interplay in Trust Games, we test a predication that for supporting cooperation, perfect information (parallel transmission) is more effective than gossip (e.g. diffusion), gossip than noisy gossip (triadic transmission) and noisy gossip than no message\cite{fonseca_will_2018}. By definition, cooperative resource exchanges lead to growth, but whether Cs benefit from it is another question. Because we program our agents to condition cooperation on reputation of the partner, Ds need Cs to perceive them as cooperative (have ‘good’ reputation) in order to receive resources. At the same time, only by defecting against cooperators do they actually acquire more resources than Cs. We compare the resources of those two groups in two information conditions by taking a mean number of resources of each group, subtracting one from the other and dividing the difference by a standard deviation of the total resource distribution for each model run. This way we achieve a measure of who gets more and by how much (relative to total dispersion). This also allows for direct comparison among differently-sized runs.

Second hypothesis focuses on how access to reputations shapes the overall resource level and its distribution. The crucial aspect of Trust Game lies within the multiplier of resources sent to trustees (3 in this model and most studies). Since cooperating with Ds is costly, it is sensible to assume that knowledge transfer will impede the money transfer – the more Cs know about their unfriendly environment, the less they will invest. At the same time, higher overall resources do not mean that Cs get a larger piece of the pie, so it is especially interesting to see if growing resource inequalities imply that the Cs are winning or not. Some studies suggest that the larger dispersion of wealth is characteristic of scenarios in which Cs outperform Ds\cite{tsvetkova_effects_2021}. We distinguish between resources of the population and those belonging to Cs and Ds respectively. We test for their differences and linear associations, as well as overall dispersion.

Third hypothesis inquires if it is true that structured game interactions (static and dynamic networks) remain safe havens for cooperation\cite{melamed_prosocial_2017}, \cite{takacs_networks_2021} paired with indirect reciprocity\cite{fuentes_ecology_2025}, when reputations are scarce (triadic gossip). To check whether this holds in this version of the Trust Game under different information conditions, we introduced a well-mixed scenario for a baseline and two scenarios where TG is played on a network.

On a finer note, previous research\cite{melamed_inequality_2022} suggests that networks are vehicles that establish and reinforce inequalities, because agents prefer playing with wealthy cooperators, even when such relationship is costly. Over time, this leads to accumulation of wealth and network centralization benefiting the few. The limitation of that work is the lack of implementation of conditional cooperation in the experimental design (players can only cooperate with everyone or not, while individual-specific decisions are made in ego-network composition stage). The present simulation has improved upon that in two ways. First, we introduce player-specific decisions (agents play with individual partners (network neighbors), cooperating with some, while defecting against others). Second, we implement game decisions as reputation-driven, where agents weight gossip information with direct experience.

\section{Results}

The following section is based on model runs with fixed amount of gossiping and fixed amount of TG rounds per run. Each simulation consists of 1000 steps representing that many gossip interactions and 10 TG rounds (plus a burn-in period of 2 rounds – see ODD). We vary the initial proportion of defectors, thresholds for cooperation, action rules and variants of TG interactions. That way we can show general tendencies, controlling for different initial environments. Agents don’t change their ‘types’ over the course of the simulation, so the composition of the population remains constant throughout the simulation. The key model indicators are the statistical characteristics of resources of groups.

The first hypothesis finds support. In parallel transmission condition Cs are, on average, 1.4 times more likely to win and get higher payoffs (t-test = 194.08; K-S = 0.20; all p$\approx$0) than in triadic gossiping, although they only allow Cs to outcompete Ds 43\% and 31\% of time respectively. Comprehensive information guides better decision-making (Cs can discriminate better).

At the same time, triadic gossiping can support cooperation, but only under very restrictive conditions, mostly driven by interplay of thresholds, action rules and TG interaction regimes (see discussion). While the cooperators’ success clearly depends on the number of available partners of type C (Fig. 1B), there is a tipping point in the model, when agents use middle regions of the reputation thresholds (Fig. 1C), at which Cs can discriminate much better, leading them to outcompete Ds on average. The smaller the threshold, the more liberal Cs are when assessing their partners, so it’s understandable that they should be exploited more on such occasions. When Cs become overly critical (high threshold value), they loose often and to a larger extent. This effect is present for all conditions (S6A), even accounting for larger proportion of Ds (S6B).

Furthermore, Sensitivity Analysis (see SI) shows that this threshold phase transition disappears when agents rely only on image (direct experience) in their decision-making for the same amount of TG rounds, but reappears, when there are ten times more TG interactions (see S7 and S8). This suggests that gossiping is about one order of magnitude more efficient in supporting cooperation than direct experience – something that hasn’t been shown before in models that combine direct and indirect reciprocity\cite{schmid_unified_2021}. In order to say how much this effect rests on the interplay between TG rounds and intensity of gossiping, we tested for different mixtures of the two (see SI for details). It turns out, that the amount of triadic gossiping does not drive the success of Cs in the shorter run (S1a), but when the triadic gossip information accumulates over time (S1b), it can improve Cs’ performance beyond direct reciprocity.

\begin{figure}[ht]
\centering
\includegraphics[width=\linewidth]{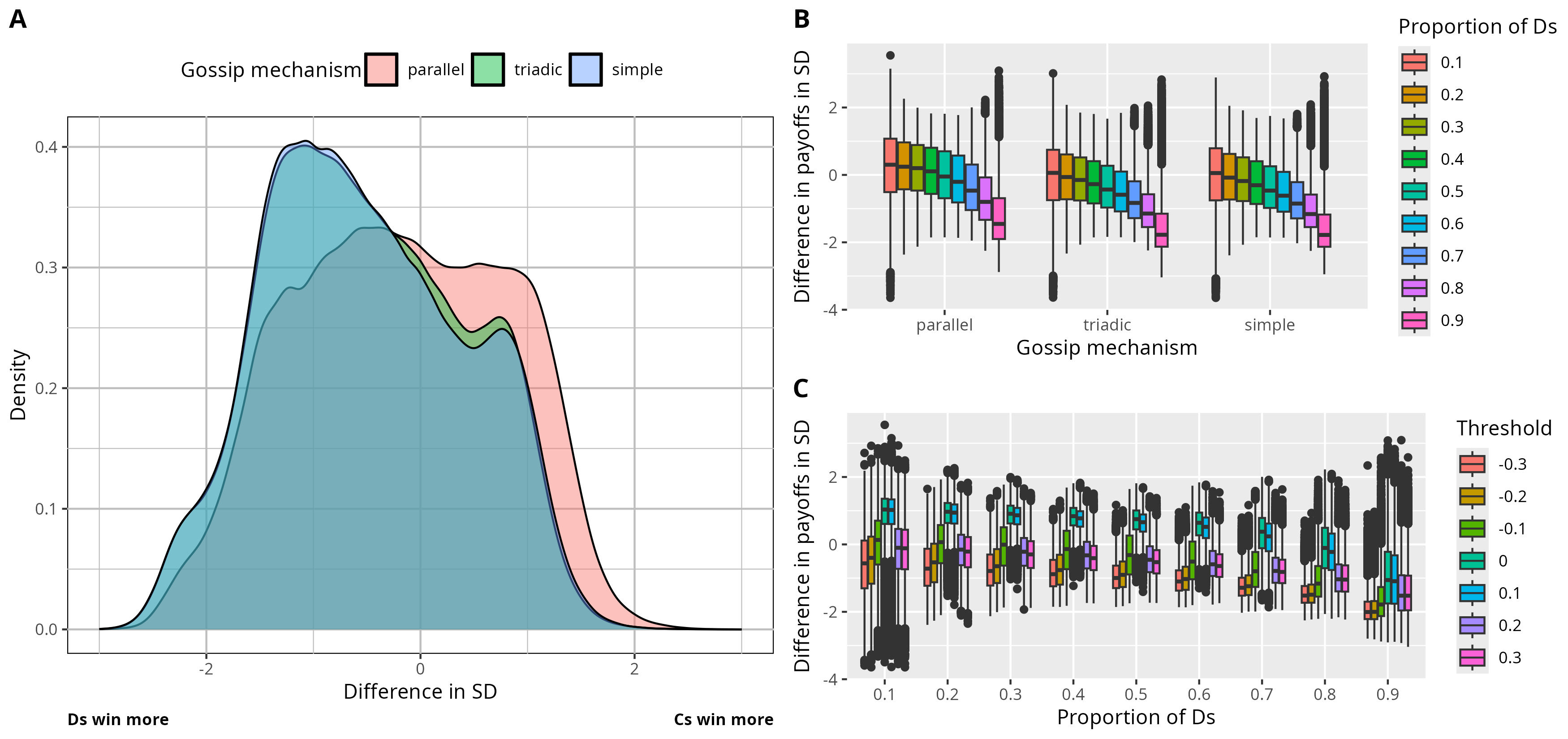}
\caption{Panel A: comparison of three density plots (Y axis) of gossip mechanisms (parallel, triadic and simple) in their support for cooperation measured by difference in mean resources between two groups divided by the standard deviation for all of the resources (X axis). Panel B: breakdown of influence of amount of Ds in the system (colors) for each gossip mechanism (X axis) measured in difference in SD (Y axis). Panel C:  difference in SD (Y axis) for reputation thresholds (colored bars) grouped for proportion of Ds in the system (X axis).}
\label{fig1}
\end{figure}

The second hypothesis concerns growth and distribution of resources. The type of information transfer does drive growth of both overall and cooperators’ resource levels (Fig. 2a). Triadic gossiping is much more likely to produce higher overall payoffs than parallel transmission (t-test= -108.53; K-S= 0.14; all p$\approx$0), so second hypothesis finds support. This same mechanism is responsible for rising resources of Cs and Ds – when playing in triadic gossip condition their resources on average grow by about 0.5 and 8 units respectively, compared to parallel. One explanation suggests that when Cs don’t know everyone’s reputation, they are more trusting and exercise their default option (cooperate with strangers) more frequently. From this we can conclude that the less Cs know, the more likely they are to invest and the higher payoffs everyone earns. Even wasteful cooperation leads to growth. However, the contrary is also true: having perfect information (compared with triadic gossip) makes one more cautious, which leads to more efficient allocation of trust, but at the cost of (s)lower growth. Without multiplication of resources from cooperative exchanges there is a ‘freezing effect’ when bad experiences and reputations accumulate, leading to distrust among agents, especially when thresholds for conditional cooperation are high (growth stagnation beyond threshold of 0.1 in Fig. 2a).

\begin{figure}[ht]
\centering
\includegraphics[width=\linewidth]{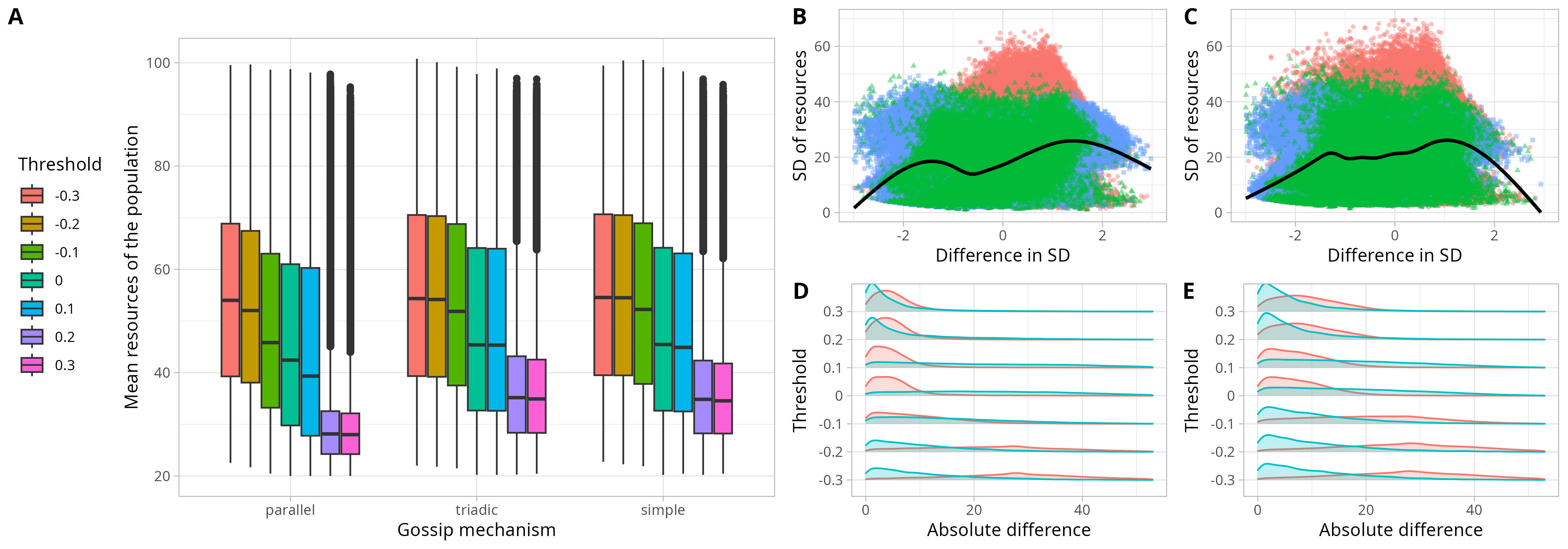}
\caption{Panel A: mean resource level of cooperators (Y axis) for different information conditions and thresholds (colored boxes). Scatterplots with fitted regression lines for parallel (B, D) and triadic gossip conditions (C, E) comparing relative difference in resources (X axis) and overall variability (SD on Y axis in B and C) for different TG interaction regimes (red: dynamic networks, green: static networks, blue: well-mixed). Histograms compare the densities of absolute difference in resources (X axis in D and E) in situations when Cs win (blue) and Ds win (red) for different thresholds (Y axis).}
\label{fig2}
\end{figure}

The story of inequalities in resources is more complex. We consider two aspects: the overall dispersion of resources (Figs. 2B and 2C) and the absolute distance between two groups’ means (Figs. 2D and 2E). As agents produce more resources, its variability (standard deviation) also grows, but that dynamic seems to be nonlinear, creating two separate regions. As Ds outcompete Cs by a larger amount (Fig. 2B and 2C left), the regression line falls, indicating a decrease in overall resource dispersion, with a much less pronounced decrease when the Cs are winning more (Fig. 2B and 2C2 right). The two peaks indicate that both groups contribute to increasing the dispersion regardless of information spread, but numerical analysis points to triadic gossip driving it further by almost 2 units. This effect is even stronger when agents play dynamic network scenario (red clouds at the top of 2B and 2C), but while they contribute most to the dispersion (almost 3 units more than other two conditions), their effect on absolute difference is small (comparable to static and 7 units less than well-mixed).

When we break down the absolute difference into gossip mechanisms (Fig. 2D and 2E) for scenarios when Cs outcompete Ds (blue) and the other way around (red), we can clearly see the pattern of triadic gossip (2E) pushing the disparity of resources into the extreme more than parallel gossip (2D). When Cs are winning, on average they contribute to the absolute difference 1.5 units less, than in D-win scenarios (Fig. S9H), but almost 5 units more to the overall dispersion (Fig. S9I). This puzzling mismatch may be caused by the fact that Ds are able to play Cs for a little more units, especially when Ds are numerous (consult bimodality of D-win contribution to both types of inequalities in S9G and S9I with more concentrated and sequential distributions for C-wins). Interestingly, when Cs win in triadic gossip, the absolute difference mismatch grows to almost 5 units, but the dispersion is lower than in parallel condition. The latter is probably driven by efficiency of dynamic networks for Cs (2B).

These results are somewhat in line with previous literature\cite{tsvetkova_effects_2021}, since we don’t allow agents to switch their type) and without it, a more realistic reputation systems contribute to growing inequalities, instead of alleviating them. To reinforce these findings, we’ve studied linear associations between standard deviation and mean resource levels (see SI). Surprisingly, when accounting for higher overall payoffs, we observed that the mean resources of Cs have a negative sign, which further corroborates that the overall dispersion of resources is reduced when Cs get richer. However, when Cs outcompete Ds, this fact highly contributes to overall dispersion (SD grows by more than 5 units), so the statistical effect of reduction in resource inequality probably comes from situations when Cs loose.

The third hypothesis examines the dynamic network condition by comparing its effectiveness in supporting cooperation with TG interactions in well-mixed and static network regimes. Results indicate (Figs. 3A-D) that partner selection implemented by dynamic networks (agents choose their own collaborators based on reputations) did produce the best situation for cooperation overall. In terms of difference between two groups, it was better than static networks (t = 99.6; K-S = 0.10; all p$\approx$0), and much better than well-mixed variant of TG (t-test = 336.6; KS = 0.33; all p$\approx$0). Dynamic networks come close to even producing sustained advantage. When it comes to information availability, parallel transmission outcompetes gossip for every action rule and TG variant (Fig. 3C and 3D). Regardless of proportion of Ds and thresholds, a particular combination of parameters of: action rule III (resembling generous TFT), static and dynamic networks allows Cs to win in most scenarios (see SI), which does not happen in well-mixed version.

\begin{figure}[ht]
\centering
\includegraphics[width=\linewidth]{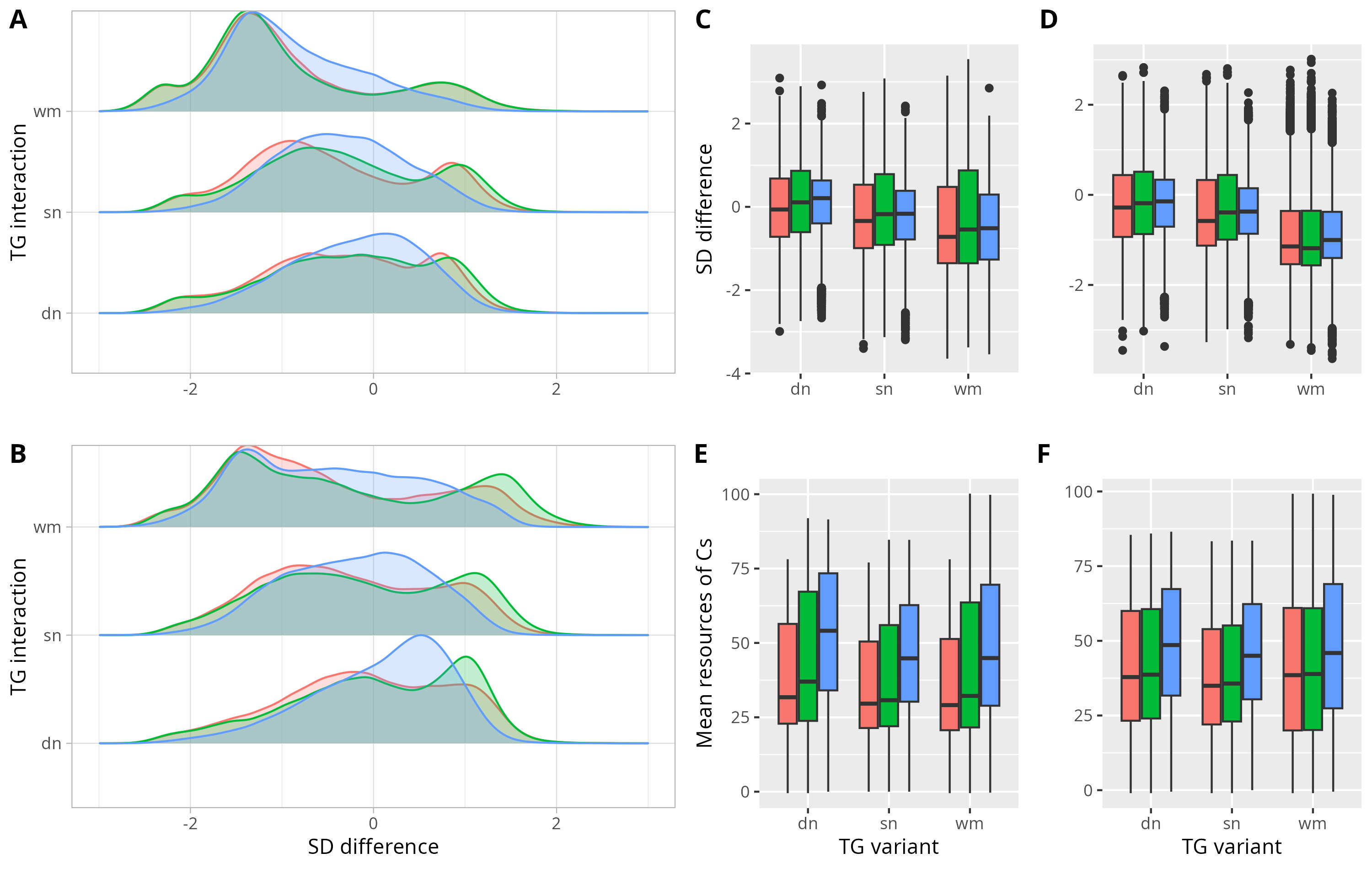}
\caption{Density (A,B) and boxplots (C-F) comparing action rules (red: 1, green: 2, blue: 3) in different TG interaction regimes (wm: well-mixed, sn: static networks, dn: dynamic networks). Panels A and B: breakdown of interactions between TG interaction and action rules for triadic (A) and parallel (B) gossip mechanisms. Panels C-F: contribution of TG interaction and action rules for triadic (D, F) and parallel (C, E) gossip mechanisms measured with relative difference (C, D) and mean resources of Cs (E, F).}
\label{fig3}
\end{figure}

Dynamic networks produce another important competitive advantage for Cs. They increase their resources by 1.5 unit more than well-mixed (t-test= 34.2; K-S= 0.06; all p$\approx$0), as well as 4 units more than static networks (t-test= 100.9; K-S= 0.08 all p$\approx$0). When we break down this into gossip mechanisms, the disparity between TG interaction regimes seems to be smaller in triadic gossip. Well-mixed and static networks produce more resources for Cs in triadic gossip condition (3F), but dynamic networks are more productive in parallel gossip. All in all, resources of Cs grow in realistic gossip more by 0.5 unit.

To corroborate these findings, we’ve run scenarios for the different time frames. It turns out, that as agents play more TG rounds in the same period (S4A), the direct experience actually benefits the well-mixed condition much more than networks. However, the static networks benefit in a similar manner (S4B) if the gossiping occurs over a longer time span. The support for higher results for Cs is even stronger in triadic gossip condition. We also manipulated the density of Trust Game network to check how sensitive our results are to sparser and denser neighborhoods. The growing number of potential partners seems to have a confusing effect on Cs in both types of networks (S5A). A similar dynamic can be observed for comparing the probability of playing exclusively with TG network neighbors (S5B): the larger it is, the better Cs get at discriminating against Ds.

\section{Discussion}

Our results show that the means by which reputations reach agents matters for sustainability of cooperation, as well as growth. When reputations are scarce (triadic gossip), cooperators are less likely to discriminate against defectors, but gather more resources in TG. Reputation is a very powerful mechanism, but adding realism to the model paints an interesting picture. When agents have full information about their peers (parallel transmission), cooperators can outcompete defectors regularly, but not most of the time, while triadic gossip only supports cooperation under restrictive conditions. Our results thus support the prediction that larger availability of reputation  increases the cooperation levels, leading them to higher relative profits.  At the same time, when cooperators base their TG decisions on this limited information, their naivety leads to growth of population resources (through the multiplier), but also positively contributes to their own average income. It seems that it is marginally more beneficial to be vulnerable (run a risk of playing against a defector perceived as someone with neutral or good reputation) in order to get richer, than to be more precise in assessment of the partner, but get stuck in stagnation.

The realism of portrayal of gossip introduced here rests on implementation of spread of gossip as a diffusion on a valued social network. Comparison with perfect information (parallel transmission) has been the focus of this paper, but we’ve also checked how this mechanism of limited information availability fares against a simple network diffusion (see SI). Surprisingly, the difference between the simple diffusion of gossip and realistic gossiping is not very large (Fig. 1A). This effect can be explained by the difference between perfect and imperfect information. It seems that the degree of this imperfection is negligible, since agents occupying local information niche on average do the same, as they would in more restrictive conditions. The real advantage comes from full information, not a little bit more, but this is not possible in the real world. These results are in line with previous research on gossip in different real settings, where cooperation can be improved\cite{fehr2019gossip}, but not as much as in a situation with perfect information\cite{sommerfeld2008multiple}.

A possible explanation for this relatively weak effect of triadic gossiping might be due to the fact that we studied it in isolation. In reality we’re entangled in multiple, co-evolving relational processes, normative and evaluative regimes competing with reputations, as well as individual interactional cues that are all very difficult to capture in a tractable computational model. If spread of reputations and subsequent rounds of TG were subjected to more ‘closure’ (i.e. the networks of communication and collaboration were not independent), they might have supported cooperation more often. At the same time, such cohesive subgroups show tendencies to create an echo effect\cite{burt_brokerage_2005}, which mostly reinforces the opinions and existing connections already present in some configuration (i.e. selection bias of what to talk about inside a support clique), which makes it hard for agents to challenge them. This is especially problematic when direct experience contradicts gossip. Our model shows this when comparing triadic and simple diffusions, as the two enter direct reciprocity regime in the long run (see S2A) – the more agents know from their own experience, the more confused they get when a lot of third-party information comes in to contradict it (as in case of simple gossip). This preliminary observation requires more studies, but it seems reasonable to claim that challenging someone’s well-formed position requires much more than simple honesty of a signal\cite{takacs_networks_2021}. At the same time, we did not introduce any status characteristics\cite{kruidhof_coevolution_2026} or network closure.

Our second contribution concerns growth and inequalities. Accumulation of resources happens when agents keep playing with each other in a cooperative manner as trustors (multiplier). This benefits both Cs and Ds, but in order for this mechanism of growth to be sustainable, agents need incentives to keep cooperating, which is at odds with Ds gaining bad reputation as they keep exploiting others. Specifically, we were interested if more information about one’s partners enhances production of the resources, and if those situations produce an advantage for cooperation. We’ve shown that less reputational information (triadic gossip) actually leads to higher payoffs, but not to a competitive advantage. Moreover, being careful in partner assessment (high cooperation threshold) leads to lower resource levels, so there seems to be a limit to more careful examination, as too much of it produces neither a relative advantage, nor growth. The same scrutiny that helps cooperators detect cheaters has an adverse effect on production of resources (S9A shows how most liberal assessments create the largest growth and vice versa). Validating one’s partners is an expensive and time-consuming endeavor and can quickly lead to stagnation (top panel in S9B shows how a restrictive threshold makes the influence of proportion of Ds diminish). The general tendency that exploitation of trust (or running the risk of it) supports growth is even more pronounced in situations when cooperators actually win, since then they create the richest populations of all by being most liberal in their partners’ assessments (S9C), especially in triadic gossip (S9J).

When inequality (both in terms of resource dispersion and the absolute difference between two groups) increases, the cooperators are more likely to drive them both. The first may be caused by their success in establishing competitive advantage (middle values of thresholds in S9G show how successful runs increase dispersion more than unsuccessful runs). The second probably comes from a liberal stance of Cs, which leads them to lose to a larger extent (S9F bottom panel). This is in line with review of experimental literature\cite{tsvetkova_effects_2021} and bears some similarity to inequalities in physical capital accumulation as a growth mechanism\cite{galor2004physical}.

At the same time, since larger inequalities may in fact indicate that cooperation is thriving, we have to consider serious implications of this result. First, though it may seem like a good idea to support the divergence between two groups,  we’ve established that Cs winning implies smaller absolute difference and larger overall dispersion (especially in dynamic networks), so this may not always maximize the benefit of cooperators. Second, because our model does not allow for agents to switch types during simulation run and we don’t condition production of new resources on existing wealth, we cannot comment on how reputations change inequalities in a dynamic setting. Such a process might work towards equality in principle (as Cs get richer, some poor Ds join them thus alleviating the inequalities between two groups), but may also have important adverse effects for both sides’ motivation to make the change. As any two groups become more unlike, the Cs may not benefit accepting anyone else (e.g. when including a poor agent will lower production output in the short run), as well as Ds may perceive their situation as a vicious circle (e.g. not having any collaborative partners leads to sucker’s payoff or reinforcing defection for self-protection). All in all, these results indicate that we should be careful in making general statements and the nature of interplay between resource growth, increasing cooperation and inequalities requires more attention. In real settings we’re very often faced with presence of established elites that have multiple incentives to keep their wealth and means of its generation restricted\cite{shennan_property_2011}. Present work sheds some light on how a reputation-based assessment already reinforces these observations (see S9C), as cooperation may by an engine of growth, but the division of its fruits becomes parochial in nature. What is good for cooperation in the intergroup competition may not be good for productivity of cooperators.

Third question concerned influence of Trust Game interaction regimes on results. Literature suggests that cooperators can thrive in dynamic network settings, if they establish clusters of cooperation – relatively homogeneous subgraphs consisting mostly or exclusively of prosocial agents\cite{takacs_networks_2021}. In this work, we’ve compared dynamic networks with static and well-mixed scenarios. Both network conditions do support cooperation more then well-mixed version, but only dynamic networks contribute to growth on par with well-mixed. This trade-off indicates that it is difficult to maximize predictability, while also maximizing development. In fact, given a prosocial orientation of some of the agents, the more open the interaction structure, the more likely their exchange is going to be productive. This further corroborates the above argument by adding a third ingredient of growth: the open selection of potential partners. Interestingly, the dynamic networks are actually better than static ones in supporting cooperation when there are many TG network neighbors (S5), so it seems that the ability to break free from an even relatively open situation (lots of potential partners) is more beneficial than verifying those partners.

There is a caveat to our implementation of dynamic networks, since we introduced ‘partner selection’ without any ‘partner acceptance’ mechanism (agent chosen by another agent to be their TG network neighbor cannot disagree), we can’t rule out that cooperators simply juggle the ‘hot potatoes’ of defection instead of creating a durable cooperation clique. Future experiments should consider this distinction, since in real life we encounter both situations: when we can opt out from a collaboration proposed by someone we know to be a burden, as well as when we must endure someone simply because they want to work with us.

In conclusion, this paper improved upon previous work in several aspects. First, we’ve build a model that introduces more realism into process of gossiping \cite{milinski_reputation_2016} and works on empirical social networks, not on stylized ones\cite{ohtsuki_reputation_2015}. Second, we’ve established that imperfect social information available through realistic gossip significantly reduces the success rate of cooperation, but even then, if cooperators show very little leniency, they can outcompete defectors on a regular basis. Third, we’ve shown that the relationships between growth, inequalities and cooperation are complex and constitute a series of trade-offs. Growing inequalities lend mixed support to cooperation, while maximizing production and reaping its benefits are difficult to achieve at the same time. This seems to be tied to emergent parochiality, which occurs on the system level in the triadic gossip condition to a much larger extent. Although cooperators are less likely to outcompete defectors, when they do so in triadic gossip condition, they become extremely productive. Fourth, the dynamic networks did support cooperation to the largest extent. This can be contributed to lack of more restrictive selection mechanism, but other implementations should be tested, as cooperators often fail to select the best partners due to lack of reputational information. We’ve built a model that combines partner selection with dyadic interactions, which improves upon previous work\cite{melamed_inequality_2022}, exposing the pitfalls of local information and local exchange. Finally, we’ve studied the interplay between indirect/direct reciprocity and gossip\cite{schmid_unified_2021}. When reputations are based on direct experiences, the system becomes dependent on the latter. However, even when scarce, gossip seems to be an effective tool in supporting cooperation, improving their decision-making by an order of magnitude.

The limitations and future directions of this work include independence of gossiping and TG rounds, lack of agent type evolution, as well as difficulty in scalability of results. First, we should add gossip-TG networks embeddedness, as in many real settings (e.g. organizations) these are dependent and agents who have good opinion about each other are likely to have a lot of positive closure in collaboration network. Second, in terms of interplay among growth, inequality and cooperation, future work should focus more on studying the dynamic aspect of switching groups and choosing partners with and without acceptance mechanism – this could relate to more opaque markets, where hosts cannot fend of parasites or providers cannot refuse consumers, as well as condition the benefit of dyadic exchanges on production capacity of both sides. Third, the model relies on empirical datasets (especially the signs of ties), so it is impossible to simply scale the population up to a large number limit without loosing the pattern of connectivity. Perhaps another type of signed data (e.g. social media or country alliances) could provide a proxy for larger populations, but reliability of introducing gossiping for big systems remains an open matter. Lastly, synchronous, as well as a common resource games should be implemented within GODS framework to assess whether results hold for double contingency and lack of dyadic interactions. Further studies are needed to disentangle remaining puzzles of reputation-based cooperation in network settings, as finding solutions to these problems helps us better address challenges of an increasingly divided world.

\section{Additional information}

Model code, datasets and ODD protocol can be found in SI. 

\bibliographystyle{naturemag}
\bibliography{GODS_refs}

\begin{thebibliography}{10}
\expandafter\ifx\csname url\endcsname\relax
  \def\url#1{\texttt{#1}}\fi
\expandafter\ifx\csname urlprefix\endcsname\relax\def\urlprefix{URL }\fi
\providecommand{\bibinfo}[2]{#2}
\providecommand{\eprint}[2][]{\url{#2}}

\bibitem{axelrod1981evolution}
\bibinfo{author}{Axelrod, R.} \& \bibinfo{author}{Hamilton, W.~D.}
\newblock \bibinfo{title}{The evolution of cooperation}.
\newblock \emph{\bibinfo{journal}{Science}} \textbf{\bibinfo{volume}{211}},
  \bibinfo{pages}{1390--1396} (\bibinfo{year}{1981}).

\bibitem{giardini_gossip_2019}
\bibinfo{author}{Giardini, F.} \& \bibinfo{author}{Wittek, R.}
\newblock \bibinfo{title}{Gossip, {Reputation}, and {Sustainable}
  {Cooperation}: {Sociological} {Foundations}}.
\newblock In \bibinfo{editor}{Giardini, F.} \& \bibinfo{editor}{Wittek, R.}
  (eds.) \emph{\bibinfo{booktitle}{The {Oxford} {Handbook} of {Gossip} and
  {Reputation}}}, \bibinfo{pages}{21--46} (\bibinfo{publisher}{Oxford
  University Press}, \bibinfo{year}{2019}).
\newblock
  \urlprefix\url{https://academic.oup.com/edited-volume/34260/chapter/290459977}.

\bibitem{milinski2019gossip}
\bibinfo{author}{Milinski, M.}
\newblock \bibinfo{title}{Gossip and reputation in social dilemmas}.
\newblock \emph{\bibinfo{journal}{The Oxford handbook of gossip and
  reputation}} \bibinfo{pages}{193--213} (\bibinfo{year}{2019}).

\bibitem{schmid_unified_2021}
\bibinfo{author}{Schmid, L.}, \bibinfo{author}{Chatterjee, K.},
  \bibinfo{author}{Hilbe, C.} \& \bibinfo{author}{Nowak, M.~A.}
\newblock \bibinfo{title}{A unified framework of direct and indirect
  reciprocity}.
\newblock \emph{\bibinfo{journal}{Nature Human Behaviour}}
  \textbf{\bibinfo{volume}{5}}, \bibinfo{pages}{1292--1302}
  (\bibinfo{year}{2021}).
\newblock \urlprefix\url{https://www.nature.com/articles/s41562-021-01114-8}.

\bibitem{szamado2021language}
\bibinfo{author}{Sz{\'a}mad{\'o}, S.}, \bibinfo{author}{Balliet, D.},
  \bibinfo{author}{Giardini, F.}, \bibinfo{author}{Power, E.} \&
  \bibinfo{author}{Tak{\'a}cs, K.}
\newblock \bibinfo{title}{The language of cooperation: reputation and honest
  signalling} (\bibinfo{year}{2021}).

\bibitem{fonseca_is_2021}
\bibinfo{author}{Fonseca, M.~A.} \& \bibinfo{author}{Peters, K.}
\newblock \bibinfo{title}{Is it costly to deceive? {People} are adept at
  detecting gossipers' lies but may not reward honesty}.
\newblock \emph{\bibinfo{journal}{Philosophical Transactions of the Royal
  Society B: Biological Sciences}} \textbf{\bibinfo{volume}{376}},
  \bibinfo{pages}{20200304} (\bibinfo{year}{2021}).
\newblock
  \urlprefix\url{https://royalsocietypublishing.org/doi/10.1098/rstb.2020.0304}.

\bibitem{righi_gossip_2022}
\bibinfo{author}{Righi, S.} \& \bibinfo{author}{Takács, K.}
\newblock \bibinfo{title}{Gossip: {Perspective} {Taking} to {Establish}
  {Cooperation}}.
\newblock \emph{\bibinfo{journal}{Dynamic Games and Applications}}
  \textbf{\bibinfo{volume}{12}}, \bibinfo{pages}{1086--1100}
  (\bibinfo{year}{2022}).
\newblock \urlprefix\url{https://link.springer.com/10.1007/s13235-022-00440-4}.

\bibitem{giardini_four_2022}
\bibinfo{author}{Giardini, F.}, \bibinfo{author}{Balliet, D.},
  \bibinfo{author}{Power, E.~A.}, \bibinfo{author}{Számadó, S.} \&
  \bibinfo{author}{Takács, K.}
\newblock \bibinfo{title}{Four {Puzzles} of {Reputation}-{Based} {Cooperation}:
  {Content}, {Process}, {Honesty}, and {Structure}}.
\newblock \emph{\bibinfo{journal}{Human Nature}} \textbf{\bibinfo{volume}{33}},
  \bibinfo{pages}{43--61} (\bibinfo{year}{2022}).
\newblock \urlprefix\url{https://link.springer.com/10.1007/s12110-021-09419-3}.

\bibitem{correia2025evolution}
\bibinfo{author}{Correia~da Fonseca, H.} \emph{et~al.}
\newblock \bibinfo{title}{Evolution of indirect reciprocity under emotion
  expression}.
\newblock \emph{\bibinfo{journal}{Scientific Reports}}
  \textbf{\bibinfo{volume}{15}}, \bibinfo{pages}{9151} (\bibinfo{year}{2025}).

\bibitem{milinski_reputation_2016}
\bibinfo{author}{Milinski, M.}
\newblock \bibinfo{title}{Reputation, a universal currency for human social
  interactions}.
\newblock \emph{\bibinfo{journal}{Philosophical Transactions of the Royal
  Society B: Biological Sciences}} \textbf{\bibinfo{volume}{371}},
  \bibinfo{pages}{20150100} (\bibinfo{year}{2016}).
\newblock
  \urlprefix\url{https://royalsocietypublishing.org/doi/10.1098/rstb.2015.0100}.

\bibitem{giardini_gossip_2021}
\bibinfo{author}{Giardini, F.}, \bibinfo{author}{Vilone, D.},
  \bibinfo{author}{Sánchez, A.} \& \bibinfo{author}{Antonioni, A.}
\newblock \bibinfo{title}{Gossip and competitive altruism support cooperation
  in a {Public} {Good} game}.
\newblock \emph{\bibinfo{journal}{Philosophical Transactions of the Royal
  Society B: Biological Sciences}} \textbf{\bibinfo{volume}{376}},
  \bibinfo{pages}{20200303} (\bibinfo{year}{2021}).
\newblock
  \urlprefix\url{https://royalsocietypublishing.org/doi/10.1098/rstb.2020.0303}.

\bibitem{takacs_networks_2021}
\bibinfo{author}{Takács, K.} \emph{et~al.}
\newblock \bibinfo{title}{Networks of reliable reputations and cooperation: a
  review}.
\newblock \emph{\bibinfo{journal}{Philosophical Transactions of the Royal
  Society B: Biological Sciences}} \textbf{\bibinfo{volume}{376}},
  \bibinfo{pages}{20200297} (\bibinfo{year}{2021}).
\newblock
  \urlprefix\url{https://royalsocietypublishing.org/doi/10.1098/rstb.2020.0297}.

\bibitem{wittek2023evolution}
\bibinfo{author}{Wittek, R.} \& \bibinfo{author}{Giardini, F.}
\newblock \emph{\bibinfo{title}{The Evolution of Reputation-based Cooperation:
  A Goal Framing Theory of Gossip}} (\bibinfo{publisher}{Cambridge University
  Press}, \bibinfo{year}{2023}).

\bibitem{okada_cooperation_2021}
\bibinfo{author}{Okada, I.}, \bibinfo{author}{Yamamoto, H.},
  \bibinfo{author}{Akiyama, E.} \& \bibinfo{author}{Toriumi, F.}
\newblock \bibinfo{title}{Cooperation in spatial public good games depends on
  the locality effects of game, adaptation, and punishment}.
\newblock \emph{\bibinfo{journal}{Scientific Reports}}
  \textbf{\bibinfo{volume}{11}}, \bibinfo{pages}{7642} (\bibinfo{year}{2021}).
\newblock \urlprefix\url{https://www.nature.com/articles/s41598-021-86668-3}.

\bibitem{uslaner_measuring_2017}
\bibinfo{author}{Bauer, P.~C.} \& \bibinfo{author}{Freitag, M.}
\newblock \emph{\bibinfo{title}{Measuring {Trust}}}, vol.~\bibinfo{volume}{1}
  (\bibinfo{publisher}{Oxford University Press}, \bibinfo{year}{2017}).
\newblock
  \urlprefix\url{https://academic.oup.com/edited-volume/34638/chapter/295110958}.

\bibitem{kollock_emergence_1994}
\bibinfo{author}{Kollock, P.}
\newblock \bibinfo{title}{The {Emergence} of {Exchange} {Structures}: {An}
  {Experimental} {Study} of {Uncertainty}, {Commitment}, and {Trust}}.
\newblock \emph{\bibinfo{journal}{American Journal of Sociology}}
  \textbf{\bibinfo{volume}{100}}, \bibinfo{pages}{313--345}
  (\bibinfo{year}{1994}).
\newblock \urlprefix\url{http://www.jstor.org/stable/2782072}.

\bibitem{coleman_foundations_2000}
\bibinfo{author}{Coleman, J.~S.}
\newblock \emph{\bibinfo{title}{Foundations of social theory}}
  (\bibinfo{publisher}{Belknap Press of Harvard Univ. Press},
  \bibinfo{address}{Cambridge, Mass.}, \bibinfo{year}{2000}),
  \bibinfo{edition}{3. print} edn.

\bibitem{van_dijk_experimental_2021}
\bibinfo{author}{Van~Dijk, E.} \& \bibinfo{author}{De~Dreu, C.~K.}
\newblock \bibinfo{title}{Experimental {Games} and {Social} {Decision}
  {Making}}.
\newblock \emph{\bibinfo{journal}{Annual Review of Psychology}}
  \textbf{\bibinfo{volume}{72}}, \bibinfo{pages}{415--438}
  (\bibinfo{year}{2021}).
\newblock
  \urlprefix\url{https://www.annualreviews.org/doi/10.1146/annurev-psych-081420-110718}.

\bibitem{berg1995trust}
\bibinfo{author}{Berg, J.}, \bibinfo{author}{Dickhaut, J.} \&
  \bibinfo{author}{McCabe, K.}
\newblock \bibinfo{title}{Trust, reciprocity, and social history}.
\newblock \emph{\bibinfo{journal}{Games and economic behavior}}
  \textbf{\bibinfo{volume}{10}}, \bibinfo{pages}{122--142}
  (\bibinfo{year}{1995}).

\bibitem{diekmann2005evolution}
\bibinfo{author}{Diekmann, A.} \& \bibinfo{author}{Przepiorka, W.}
\newblock \bibinfo{title}{The evolution of trust and reputation: results from
  simulation experiments}.
\newblock In \emph{\bibinfo{booktitle}{Third ESSA Conference}},
  \bibinfo{pages}{1--7} (\bibinfo{year}{2005}).

\bibitem{kumar_evolution_2020}
\bibinfo{author}{Kumar, A.}, \bibinfo{author}{Capraro, V.} \&
  \bibinfo{author}{Perc, M.}
\newblock \bibinfo{title}{The evolution of trust and trustworthiness}.
\newblock \emph{\bibinfo{journal}{Journal of The Royal Society Interface}}
  \textbf{\bibinfo{volume}{17}}, \bibinfo{pages}{20200491}
  (\bibinfo{year}{2020}).
\newblock
  \urlprefix\url{https://royalsocietypublishing.org/doi/10.1098/rsif.2020.0491}.

\bibitem{johnson_trust_2011}
\bibinfo{author}{Johnson, N.~D.} \& \bibinfo{author}{Mislin, A.~A.}
\newblock \bibinfo{title}{Trust games: {A} meta-analysis}.
\newblock \emph{\bibinfo{journal}{Journal of Economic Psychology}}
  \textbf{\bibinfo{volume}{32}}, \bibinfo{pages}{865--889}
  (\bibinfo{year}{2011}).
\newblock
  \urlprefix\url{https://linkinghub.elsevier.com/retrieve/pii/S0167487011000869}.

\bibitem{milinski_reputation_2002}
\bibinfo{author}{Milinski, M.}, \bibinfo{author}{Semmann, D.} \&
  \bibinfo{author}{Krambeck, H.-J.}
\newblock \bibinfo{title}{Reputation helps solve the ‘tragedy of the
  commons’}.
\newblock \emph{\bibinfo{journal}{Nature}} \textbf{\bibinfo{volume}{415}},
  \bibinfo{pages}{424--426} (\bibinfo{year}{2002}).
\newblock \urlprefix\url{https://www.nature.com/articles/415424a}.

\bibitem{chen_network-based_2015}
\bibinfo{author}{Chen, S.-H.}, \bibinfo{author}{Chie, B.-T.} \&
  \bibinfo{author}{Zhang, T.}
\newblock \bibinfo{title}{Network-{Based} {Trust} {Games}: {An} {Agent}-{Based}
  {Model}}.
\newblock \emph{\bibinfo{journal}{Journal of Artificial Societies and Social
  Simulation}} \textbf{\bibinfo{volume}{18}}, \bibinfo{pages}{5}
  (\bibinfo{year}{2015}).
\newblock \urlprefix\url{http://jasss.soc.surrey.ac.uk/18/3/5.html}.

\bibitem{greif_contract_1993}
\bibinfo{author}{Greif, A.}
\newblock \bibinfo{title}{Contract {Enforceability} and {Economic}
  {Institutions} in {Early} {Trade}: {The} {Maghribi} {Traders}' {Coalition}}.
\newblock \emph{\bibinfo{journal}{The American Economic Review}}
  \textbf{\bibinfo{volume}{83}}, \bibinfo{pages}{525--548}
  (\bibinfo{year}{1993}).
\newblock \urlprefix\url{http://www.jstor.org/stable/2117532}.

\bibitem{diekmann_trust_2019}
\bibinfo{author}{Diekmann, A.} \& \bibinfo{author}{Przepiorka, W.}
\newblock \bibinfo{title}{Trust and {Reputation} in {Markets}}.
\newblock In \bibinfo{editor}{Giardini, F.} \& \bibinfo{editor}{Wittek, R.}
  (eds.) \emph{\bibinfo{booktitle}{The {Oxford} {Handbook} of {Gossip} and
  {Reputation}}}, \bibinfo{pages}{381--400} (\bibinfo{publisher}{Oxford
  University Press}, \bibinfo{year}{2019}).
\newblock
  \urlprefix\url{https://academic.oup.com/edited-volume/34260/chapter/290471304}.

\bibitem{jiao_reputation_2021}
\bibinfo{author}{Jiao, R.}, \bibinfo{author}{Przepiorka, W.} \&
  \bibinfo{author}{Buskens, V.}
\newblock \bibinfo{title}{Reputation effects in peer-to-peer online markets:
  {A} meta-analysis}.
\newblock \emph{\bibinfo{journal}{Social Science Research}}
  \textbf{\bibinfo{volume}{95}}, \bibinfo{pages}{102522}
  (\bibinfo{year}{2021}).
\newblock
  \urlprefix\url{https://linkinghub.elsevier.com/retrieve/pii/S0049089X20301204}.

\bibitem{burt_brokerage_2005}
\bibinfo{author}{Burt, R.~S.}
\newblock \emph{\bibinfo{title}{Brokerage and closure: an introduction to
  social capital}}.
\newblock Clarendon lectures in management studies (\bibinfo{publisher}{Oxford
  University Press}, \bibinfo{address}{Oxford}, \bibinfo{year}{2005}).

\bibitem{boero_reputational_2009}
\bibinfo{author}{Boero, R.}, \bibinfo{author}{Bravo, G.},
  \bibinfo{author}{Castellani, M.} \& \bibinfo{author}{Squazzoni, F.}
\newblock \bibinfo{title}{Reputational cues in repeated trust games}.
\newblock \emph{\bibinfo{journal}{The Journal of Socio-Economics}}
  \textbf{\bibinfo{volume}{38}}, \bibinfo{pages}{871--877}
  (\bibinfo{year}{2009}).
\newblock
  \urlprefix\url{https://linkinghub.elsevier.com/retrieve/pii/S1053535709000675}.

\bibitem{feinberg_virtues_2012}
\bibinfo{author}{Feinberg, M.}, \bibinfo{author}{Willer, R.},
  \bibinfo{author}{Stellar, J.} \& \bibinfo{author}{Keltner, D.}
\newblock \bibinfo{title}{The virtues of gossip: {Reputational} information
  sharing as prosocial behavior.}
\newblock \emph{\bibinfo{journal}{Journal of Personality and Social
  Psychology}} \textbf{\bibinfo{volume}{102}}, \bibinfo{pages}{1015--1030}
  (\bibinfo{year}{2012}).
\newblock \urlprefix\url{https://doi.apa.org/doi/10.1037/a0026650}.

\bibitem{fehr2019gossip}
\bibinfo{author}{Fehr, D.} \& \bibinfo{author}{Sutter, M.}
\newblock \bibinfo{title}{Gossip and the efficiency of interactions}.
\newblock \emph{\bibinfo{journal}{Games and Economic Behavior}}
  \textbf{\bibinfo{volume}{113}}, \bibinfo{pages}{448--460}
  (\bibinfo{year}{2019}).

\bibitem{fonseca_will_2018}
\bibinfo{author}{Fonseca, M.~A.} \& \bibinfo{author}{Peters, K.}
\newblock \bibinfo{title}{Will any gossip do? {Gossip} does not need to be
  perfectly accurate to promote trust}.
\newblock \emph{\bibinfo{journal}{Games and Economic Behavior}}
  \textbf{\bibinfo{volume}{107}}, \bibinfo{pages}{253--281}
  (\bibinfo{year}{2018}).
\newblock
  \urlprefix\url{https://linkinghub.elsevier.com/retrieve/pii/S0899825617301707}.

\bibitem{peters_truth_2020}
\bibinfo{author}{Peters, K.} \& \bibinfo{author}{Fonseca, M.~A.}
\newblock \bibinfo{title}{Truth, {Lies}, and {Gossip}}.
\newblock \emph{\bibinfo{journal}{Psychological Science}}
  \textbf{\bibinfo{volume}{31}}, \bibinfo{pages}{702--714}
  (\bibinfo{year}{2020}).
\newblock
  \urlprefix\url{https://journals.sagepub.com/doi/10.1177/0956797620916708}.

\bibitem{feinberg_gossip_2014}
\bibinfo{author}{Feinberg, M.}, \bibinfo{author}{Willer, R.} \&
  \bibinfo{author}{Schultz, M.}
\newblock \bibinfo{title}{Gossip and ostracism promote cooperation in groups}.
\newblock \emph{\bibinfo{journal}{Psychological Science}}
  \textbf{\bibinfo{volume}{25}}, \bibinfo{pages}{656--664}
  (\bibinfo{year}{2014}).
\newblock
  \urlprefix\url{https://journals.sagepub.com/doi/10.1177/0956797613510184}.

\bibitem{gallo_effects_2015}
\bibinfo{author}{Gallo, E.} \& \bibinfo{author}{Yan, C.}
\newblock \bibinfo{title}{The effects of reputational and social knowledge on
  cooperation}.
\newblock \emph{\bibinfo{journal}{Proceedings of the National Academy of
  Sciences}} \textbf{\bibinfo{volume}{112}}, \bibinfo{pages}{3647--3652}
  (\bibinfo{year}{2015}).
\newblock \urlprefix\url{https://pnas.org/doi/full/10.1073/pnas.1415883112}.

\bibitem{nee2023theory}
\bibinfo{author}{Nee, V.}, \bibinfo{author}{Wang, S.} \& \bibinfo{author}{Macy,
  M.~W.}
\newblock \bibinfo{title}{A theory of emergence: knowledge, rewiring and
  innovation}.
\newblock \emph{\bibinfo{journal}{Social Science Research}}
  \textbf{\bibinfo{volume}{111}}, \bibinfo{pages}{102851}
  (\bibinfo{year}{2023}).

\bibitem{emler1994gossip}
\bibinfo{author}{Emler, N.}
\newblock \bibinfo{title}{Gossip, reputation, and social adaptation.}
  (\bibinfo{year}{1994}).

\bibitem{dunbar_gossip_2004}
\bibinfo{author}{Dunbar, R. I.~M.}
\newblock \bibinfo{title}{Gossip in {Evolutionary} {Perspective}}.
\newblock \emph{\bibinfo{journal}{Review of General Psychology}}
  \textbf{\bibinfo{volume}{8}}, \bibinfo{pages}{100--110}
  (\bibinfo{year}{2004}).
\newblock
  \urlprefix\url{https://journals.sagepub.com/doi/10.1037/1089-2680.8.2.100}.

\bibitem{ohtsuki_reputation_2015}
\bibinfo{author}{Ohtsuki, H.}, \bibinfo{author}{Iwasa, Y.} \&
  \bibinfo{author}{Nowak, M.~A.}
\newblock \bibinfo{title}{Reputation {Effects} in {Public} and {Private}
  {Interactions}}.
\newblock \emph{\bibinfo{journal}{PLOS Computational Biology}}
  \textbf{\bibinfo{volume}{11}}, \bibinfo{pages}{e1004527}
  (\bibinfo{year}{2015}).
\newblock \urlprefix\url{https://dx.plos.org/10.1371/journal.pcbi.1004527}.

\bibitem{yucel2021being}
\bibinfo{author}{Yucel, M.}, \bibinfo{author}{Sjobeck, G.~R.},
  \bibinfo{author}{Glass, R.} \& \bibinfo{author}{Rottman, J.}
\newblock \bibinfo{title}{Being in the know: Social network analysis of gossip
  and friendship on a college campus}.
\newblock \emph{\bibinfo{journal}{Human Nature}} \textbf{\bibinfo{volume}{32}},
  \bibinfo{pages}{603--621} (\bibinfo{year}{2021}).

\bibitem{dores2021gossip}
\bibinfo{author}{Dores~Cruz, T.~D.} \emph{et~al.}
\newblock \bibinfo{title}{Gossip and reputation in everyday life}.
\newblock \emph{\bibinfo{journal}{Philosophical Transactions of the Royal
  Society B}} \textbf{\bibinfo{volume}{376}}, \bibinfo{pages}{20200301}
  (\bibinfo{year}{2021}).

\bibitem{robbins_who_2020}
\bibinfo{author}{Robbins, M.~L.} \& \bibinfo{author}{Karan, A.}
\newblock \bibinfo{title}{Who {Gossips} and {How} in {Everyday} {Life}?}
\newblock \emph{\bibinfo{journal}{Social Psychological and Personality
  Science}} \textbf{\bibinfo{volume}{11}}, \bibinfo{pages}{185--195}
  (\bibinfo{year}{2020}).
\newblock
  \urlprefix\url{https://journals.sagepub.com/doi/10.1177/1948550619837000}.

\bibitem{martinescu_what_2022}
\bibinfo{author}{Martinescu, E.}, \bibinfo{author}{Peters, K.} \&
  \bibinfo{author}{Beersma, B.}
\newblock \bibinfo{title}{What {Do} {We} {Talk} {About} {When} {We} {Talk}
  {About} {Others}? {Evidence} for the {Primacy} of the {Horizontal}
  {Dimension} of {Social} {Evaluation} in {Workplace} {Gossip}}.
\newblock \emph{\bibinfo{journal}{International Review of Social Psychology}}
  \textbf{\bibinfo{volume}{35}}, \bibinfo{pages}{13} (\bibinfo{year}{2022}).
\newblock \urlprefix\url{https://www.rips-irsp.com/article/10.5334/irsp.687/}.

\bibitem{estevez_brokering_2022}
\bibinfo{author}{Estévez, J.~L.} \& \bibinfo{author}{Takács, K.}
\newblock \bibinfo{title}{Brokering or {Sitting} {Between} {Two} {Chairs}? {A}
  {Group} {Perspective} on {Workplace} {Gossip}}.
\newblock \emph{\bibinfo{journal}{Frontiers in Psychology}}
  \textbf{\bibinfo{volume}{13}}, \bibinfo{pages}{815383}
  (\bibinfo{year}{2022}).
\newblock
  \urlprefix\url{https://www.frontiersin.org/articles/10.3389/fpsyg.2022.815383/full}.

\bibitem{giardini_evolution_2016}
\bibinfo{author}{Giardini, F.} \& \bibinfo{author}{Vilone, D.}
\newblock \bibinfo{title}{Evolution of gossip-based indirect reciprocity on a
  bipartite network}.
\newblock \emph{\bibinfo{journal}{Scientific Reports}}
  \textbf{\bibinfo{volume}{6}}, \bibinfo{pages}{37931} (\bibinfo{year}{2016}).
\newblock \urlprefix\url{https://www.nature.com/articles/srep37931}.

\bibitem{samu_scarce_2020}
\bibinfo{author}{Samu, F.}, \bibinfo{author}{Számadó, S.} \&
  \bibinfo{author}{Takács, K.}
\newblock \bibinfo{title}{Scarce and directly beneficial reputations support
  cooperation}.
\newblock \emph{\bibinfo{journal}{Scientific Reports}}
  \textbf{\bibinfo{volume}{10}}, \bibinfo{pages}{11486} (\bibinfo{year}{2020}).
\newblock \urlprefix\url{https://www.nature.com/articles/s41598-020-68123-x}.

\bibitem{samu_evaluating_2021}
\bibinfo{author}{Samu, F.} \& \bibinfo{author}{Takács, K.}
\newblock \bibinfo{title}{Evaluating mechanisms that could support credible
  reputations and cooperation: cross-checking and social bonding}.
\newblock \emph{\bibinfo{journal}{Philosophical Transactions of the Royal
  Society B: Biological Sciences}} \textbf{\bibinfo{volume}{376}},
  \bibinfo{pages}{20200302} (\bibinfo{year}{2021}).
\newblock
  \urlprefix\url{https://royalsocietypublishing.org/doi/10.1098/rstb.2020.0302}.

\bibitem{estevez_more_2022}
\bibinfo{author}{Estévez, J.~L.}, \bibinfo{author}{Kisfalusi, D.} \&
  \bibinfo{author}{Takács, K.}
\newblock \bibinfo{title}{More than one’s negative ties: {The} role of
  friends’ antipathies in high school gossip}.
\newblock \emph{\bibinfo{journal}{Social Networks}}
  \textbf{\bibinfo{volume}{70}}, \bibinfo{pages}{77--89}
  (\bibinfo{year}{2022}).
\newblock
  \urlprefix\url{https://linkinghub.elsevier.com/retrieve/pii/S0378873321001064}.

\bibitem{grosser2010social}
\bibinfo{author}{Grosser, T.~J.}, \bibinfo{author}{Lopez-Kidwell, V.} \&
  \bibinfo{author}{Labianca, G.}
\newblock \bibinfo{title}{A social network analysis of positive and negative
  gossip in organizational life}.
\newblock \emph{\bibinfo{journal}{Group \& organization management}}
  \textbf{\bibinfo{volume}{35}}, \bibinfo{pages}{177--212}
  (\bibinfo{year}{2010}).

\bibitem{ellwardt_co-evolution_2012}
\bibinfo{author}{Ellwardt, L.}, \bibinfo{author}{Steglich, C.} \&
  \bibinfo{author}{Wittek, R.}
\newblock \bibinfo{title}{The co-evolution of gossip and friendship in
  workplace social networks}.
\newblock \emph{\bibinfo{journal}{Social Networks}}
  \textbf{\bibinfo{volume}{34}}, \bibinfo{pages}{623--633}
  (\bibinfo{year}{2012}).
\newblock
  \urlprefix\url{https://linkinghub.elsevier.com/retrieve/pii/S0378873312000445}.

\bibitem{giardini_silence_2019}
\bibinfo{author}{Giardini, F.} \& \bibinfo{author}{Wittek, R. P.~M.}
\newblock \bibinfo{title}{Silence {Is} {Golden}. {Six} {Reasons} {Inhibiting}
  the {Spread} of {Third}-{Party} {Gossip}}.
\newblock \emph{\bibinfo{journal}{Frontiers in Psychology}}
  \textbf{\bibinfo{volume}{10}}, \bibinfo{pages}{1120} (\bibinfo{year}{2019}).
\newblock
  \urlprefix\url{https://www.frontiersin.org/article/10.3389/fpsyg.2019.01120/full}.

\bibitem{borgonovo_sensitivity_2022}
\bibinfo{author}{Borgonovo, E.}, \bibinfo{author}{Pangallo, M.},
  \bibinfo{author}{Rivkin, J.}, \bibinfo{author}{Rizzo, L.} \&
  \bibinfo{author}{Siggelkow, N.}
\newblock \bibinfo{title}{Sensitivity analysis of agent-based models: a new
  protocol}.
\newblock \emph{\bibinfo{journal}{Computational and Mathematical Organization
  Theory}} \textbf{\bibinfo{volume}{28}}, \bibinfo{pages}{52--94}
  (\bibinfo{year}{2022}).
\newblock \urlprefix\url{https://link.springer.com/10.1007/s10588-021-09358-5}.

\bibitem{jager2017enhancing}
\bibinfo{author}{Jager, W.}
\newblock \bibinfo{title}{Enhancing the realism of simulation (eros): On
  implementing and developing psychological theory in social simulation}.
\newblock \emph{\bibinfo{journal}{Jasss-The journal of artificial societies and
  social simulation}} \textbf{\bibinfo{volume}{20}}, \bibinfo{pages}{14}
  (\bibinfo{year}{2017}).

\bibitem{masuda_coevolution_2012}
\bibinfo{author}{Masuda, N.} \& \bibinfo{author}{Nakamura, M.}
\newblock \bibinfo{title}{Coevolution of {Trustful} {Buyers} and {Cooperative}
  {Sellers} in the {Trust} {Game}}.
\newblock \emph{\bibinfo{journal}{PLoS ONE}} \textbf{\bibinfo{volume}{7}},
  \bibinfo{pages}{e44169} (\bibinfo{year}{2012}).
\newblock \urlprefix\url{https://dx.plos.org/10.1371/journal.pone.0044169}.

\bibitem{flache_models_2017}
\bibinfo{author}{Flache, A.} \emph{et~al.}
\newblock \bibinfo{title}{Models of {Social} {Influence}: {Towards} the {Next}
  {Frontiers}}.
\newblock \emph{\bibinfo{journal}{Journal of Artificial Societies and Social
  Simulation}} \textbf{\bibinfo{volume}{20}}, \bibinfo{pages}{2}
  (\bibinfo{year}{2017}).
\newblock \urlprefix\url{http://jasss.soc.surrey.ac.uk/20/4/2.html}.

\bibitem{galor2004physical}
\bibinfo{author}{Galor, O.} \& \bibinfo{author}{Moav, O.}
\newblock \bibinfo{title}{From physical to human capital accumulation:
  Inequality and the process of development}.
\newblock \emph{\bibinfo{journal}{The review of economic studies}}
  \textbf{\bibinfo{volume}{71}}, \bibinfo{pages}{1001--1026}
  (\bibinfo{year}{2004}).

\bibitem{melamed_inequality_2022}
\bibinfo{author}{Melamed, D.}, \bibinfo{author}{Simpson, B.},
  \bibinfo{author}{Montgomery, B.} \& \bibinfo{author}{Patel, V.}
\newblock \bibinfo{title}{Inequality and cooperation in social networks}.
\newblock \emph{\bibinfo{journal}{Scientific Reports}}
  \textbf{\bibinfo{volume}{12}}, \bibinfo{pages}{6789} (\bibinfo{year}{2022}).
\newblock \urlprefix\url{https://www.nature.com/articles/s41598-022-10733-8}.

\bibitem{tsvetkova_effects_2021}
\bibinfo{author}{Tsvetkova, M.}
\newblock \bibinfo{title}{The effects of reputation on inequality in network
  cooperation games}.
\newblock \emph{\bibinfo{journal}{Philosophical Transactions of the Royal
  Society B: Biological Sciences}} \textbf{\bibinfo{volume}{376}},
  \bibinfo{pages}{20200299} (\bibinfo{year}{2021}).
\newblock
  \urlprefix\url{https://royalsocietypublishing.org/doi/10.1098/rstb.2020.0299}.

\bibitem{melamed_prosocial_2017}
\bibinfo{author}{Melamed, D.}, \bibinfo{author}{Simpson, B.} \&
  \bibinfo{author}{Harrell, A.}
\newblock \bibinfo{title}{Prosocial orientation alters network dynamics and
  fosters cooperation}.
\newblock \emph{\bibinfo{journal}{Scientific Reports}}
  \textbf{\bibinfo{volume}{7}}, \bibinfo{pages}{357} (\bibinfo{year}{2017}).

\bibitem{fuentes_ecology_2025}
\bibinfo{author}{Fuentes, M.}
\newblock \bibinfo{title}{Ecology and the evolution of cooperation by partner
  choice and reciprocity}.
\newblock \emph{\bibinfo{journal}{Scientific Reports}}
  \textbf{\bibinfo{volume}{15}}, \bibinfo{pages}{4613} (\bibinfo{year}{2025}).
\newblock \urlprefix\url{https://www.nature.com/articles/s41598-025-87984-8}.

\bibitem{sommerfeld2008multiple}
\bibinfo{author}{Sommerfeld, R.~D.}, \bibinfo{author}{Krambeck, H.-J.} \&
  \bibinfo{author}{Milinski, M.}
\newblock \bibinfo{title}{Multiple gossip statements and their effect on
  reputation and trustworthiness}.
\newblock \emph{\bibinfo{journal}{Proceedings of the Royal Society B:
  Biological Sciences}} \textbf{\bibinfo{volume}{275}},
  \bibinfo{pages}{2529--2536} (\bibinfo{year}{2008}).

\bibitem{kruidhof_coevolution_2026}
\bibinfo{author}{Kruidhof, E.}, \bibinfo{author}{Corten, R.},
  \bibinfo{author}{Ellwardt, L.} \& \bibinfo{author}{Wittek, R.}
\newblock \bibinfo{title}{The co-evolution of informal social status and gossip
  in workplace social networks}.
\newblock \emph{\bibinfo{journal}{Social Networks}}
  \textbf{\bibinfo{volume}{84}}, \bibinfo{pages}{9--26} (\bibinfo{year}{2026}).
\newblock
  \urlprefix\url{https://linkinghub.elsevier.com/retrieve/pii/S0378873325000474}.

\bibitem{shennan_property_2011}
\bibinfo{author}{Shennan, S.}
\newblock \bibinfo{title}{Property and wealth inequality as cultural niche
  construction}.
\newblock \emph{\bibinfo{journal}{Philosophical Transactions of the Royal
  Society B: Biological Sciences}} \textbf{\bibinfo{volume}{366}},
  \bibinfo{pages}{918--926} (\bibinfo{year}{2011}).
\newblock
  \urlprefix\url{https://royalsocietypublishing.org/doi/10.1098/rstb.2010.0309}.
\newblock \bibinfo{note}{Publisher: The Royal Society}.

\end{thebibliography}

\end{document}